\documentclass[aps,prl,twocolumn,superscriptaddress,showpacs,preprintnumbers]{revtex4-1}
\usepackage{graphicx }
\usepackage{dcolumn}
\usepackage{bm}
\usepackage{braket}
\usepackage{amsmath}
\usepackage[caption=false]{subfig} 
\usepackage{setspace} 
\usepackage{tabularx}
\usepackage{color}
\usepackage{epstopdf}

\begin{document}


\title{Bandwidth controlled insulator-metal transition in BaFe$_2$S$_3$: A M\"ossbauer study under pressure. 
}
\author{Philipp Materne}
\altaffiliation[]{philipp.materne@gmx.de, phm@anl.gov}
\affiliation{Argonne National Laboratory, Lemont, IL 60439, USA}
\author{Wenli Bi}
\affiliation{Argonne National Laboratory, Lemont, IL 60439, USA}
\affiliation{Department of Geology, University of Illinois at Urbana-Champaign, Urbana, Illinois 61801, USA}
\author{Jiyong Zhao}
\author{Michael Yu Hu}
\affiliation{Argonne National Laboratory, Lemont, IL 60439, USA}
\author{Maria Lourdes Amig\'o}
\author{Silvia Seiro}
\author{Saicharan Aswartham}
\affiliation{Leibniz Institute for Solid State and Materials Research (IFW) Dresden, D-01069, Germany}
\author{Bernd B\"uchner}
\affiliation{Leibniz Institute for Solid State and Materials Research (IFW) Dresden, D-01069, Germany}
\affiliation{Institute of Solid State and Materials Physics, TU Dresden, D-01069 Dresden, Germany}
\author{Esen Ercan Alp}
\affiliation{Argonne National Laboratory, Lemont, IL 60439, USA}

\newcommand{\bhf}{$B_{\mathrm{hf}}$}
\newcommand{\tn}{$T_{\mathrm{N}}$}
\newcommand{\pc}{$p_c$}


\date{\today}

\begin{abstract}
BaFe$_2$S$_3$ is a quasi one-dimensional Mott insulator that orders antiferromagnetically below 117(5)\,K.
The application of pressure induces a transition to a metallic state, and superconductivity emerges.
The evolution of the magnetic behavior on increasing pressure has up to now been either studied indirectly by means of transport measurements, or by using local magnetic probes only in the low pressure region.
Here, we investigate the magnetic properties of  BaFe$_2$S$_3$  up to 9.9\,GPa by means of synchrotron $^{57}$Fe M\"ossbauer spectroscopy experiments, providing the first local magnetic phase diagram.
The magnetic ordering temperature increases up to 185(5) K at 7.5\,GPa, and is fully suppressed at 9.9\,GPa.
The low-temperature magnetic hyperfine field is continuously reduced from 12.9 to 10.3\,T between 1.4 and 9.1\,GPa, followed by a sudden drop to zero at 9.9\,GPa indicating a first-order phase transition.
The pressure dependence of the magnetic order in BaFe$_2$S$_3$ can be qualitatively explained by a combination of a bandwidth-controlled insulator-metal transition as well as a pressure enhanced exchange interaction between Fe-atoms and Fe 3\textit{d}-S 3\textit{p} hybridization.

\end{abstract}

\maketitle

Since the discovery of superconductivity in LaFeAsO$_{1-x}$F$_x$ the Fe-based superconductors have been heavily studied \cite{ja800073m}.
LaFeAsO, BaFe$_2$As$_2$, and their isostructural equivalents are semi-metals showing a spin-densitiy wave order below the magnetic phase transition temperature \tn.
Both compounds share the same structural motif: a two-dimensional Fe-As layer with a planar Fe square-lattice \cite{00018732.2010.513480}.
Superconductivity (SC) can be induced in both compounds by applying hydrostatic pressure or chemical substitution \cite{ja800073m, 0953-8984-21-1-012208, nmat2397}.
The discovery of pressure-induced SC in BaFe$_2$S$_3$ added a new class to the Fe-based SC \cite{nmat4351,PhysRevLett.115.246402}.
BaFe$_2$S$_3$ is different in two aspects: it i a Mott insulator at ambient conditions and its basic structural motif are quasi one-dimensional ladders of edge-sharing FeS$_4$ tetrahedra \cite{HONG197293,cm0000346,nmat4351}.
The compound crystallizes in the orthorhombic \textit{Cmcm} space group and orders antiferromagnetically (AFM) with a stripe-type arrangement below $\sim$ 125 K \cite{HONG197293, cm0000346,nmat4351}.
It was found that nesting can be excluded as a mechanism for the AFM order, in contrast to the FeAs-based compounds \cite{2018arXiv180300282P}.
At the critical pressure \textit{p}$_c$ $\sim$ 10\,GPa an insulator-metal transition occurs and a non-magnetic SC phase with transition temperatures up to 24\,K at 12 -- 13\,GPa was observed \cite{nmat4351,PhysRevLett.115.246402}.

So far, the evolution of magnetic order with pressure over a large pressure range has only been studied indirectly by means of resistivity measurements ~\cite{PhysRevLett.115.246402}, while investigations with magnetic local probes have only been reported up to 2.6\,GPa \cite{PhysRevLett.117.047003,2018arXiv180710703Z}.
In order to fill this knowledge gap we performed synchrotron $^{57}$Fe M\"ossbauer spectroscopy (SMS) experiments on BaFe$_2$S$_3$ single crystals as a function of temperature and pressure up to 9.9\,{}GPa.
SMS allows to independently measure the magnetic volume fraction and the magnetic hyperfine field, which is a measure for the magnetic moment, and thus the magnetic phase diagram can be explored with high precision.
We found that the pressure dependence of the low-temperature magnetic hyperfine field and the magnetic phase transition temperature are consistent with a combination of a bandwidth-controlled insulator-metal transition as well as a pressure enhanced exchange interaction between Fe-atoms and Fe 3\textit{d}-S 3\textit{p} hybridization \cite{Mott1956, mott1990metal,BLOCH1966881}.
The critical pressure $p_c$ of the Mott transition is $\sim$ 9.9\,GPa.

\begin{figure}[htbp]
	\centering
		\includegraphics[width=0.9\columnwidth]{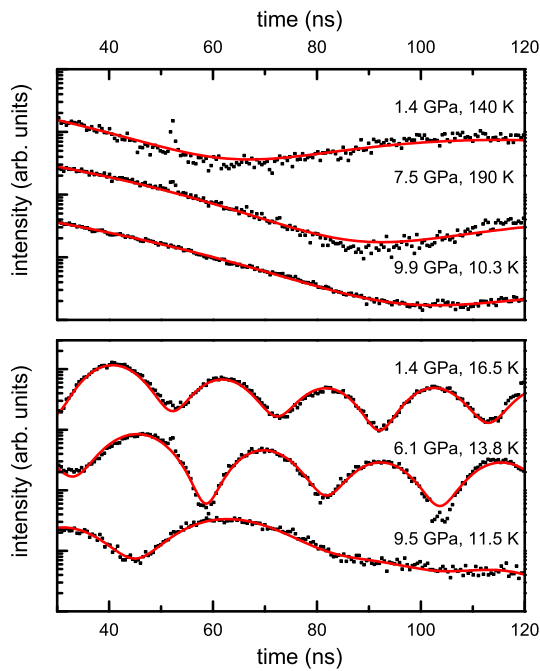}
	\caption{Synchrotron M\"ossbauer spectra at representative temperatures and pressures in the paramagnetic (top) and magnetically ordered phase (bottom). The outliers at $\sim$ 52.5\,ns are due to an impurity in the synchrotron bunches. Red lines are fits to the data.}
	\label{fig:spectra}
\end{figure}

Single crystals were grown from BaS, Fe, and S powders in a glassy carbon crucible sealed in an evacuated quartz ampule. The ampule was placed in a box furnace, heated up to 1100\,$^{\circ}$C, and cooled slowly to 750\,$^{\circ}$C. 
BaFe$_2$S$_3$ mm-sized single crystals with needle morphology were then mechanically extracted from the crucible. 
As-grown crystals were characterized by powder x-ray measurements and by energy-dispersive x-ray spectroscopy for compositional analysis. 

The SMS experiments were conducted at the beamline 3ID-B of the Advanced Photon Source, Argonne National Laboratory, USA.
The measurements were conducted between 10 and 200\,K and 1.4 and 9.9\,GPa using a special He-flow cryostat and a miniature diamond anvil cell \cite{Bihf5283,1.4999787}.
Daphne oil 7575 was used as the pressure transmitting medium to ensure quasi-hydrostatic pressure conditions.
The pressure was measured \textit{in situ} via ruby fluorescence with an uncertainty of 0.1\,GPa.
The beam size was 15$\times$20\;{}$\mu$m$^2$ (FWHM).
The SMS spectra were analyzed using CONUSS \cite{CONUSS}.

SMS spectra for representative temperatures and pressures are shown in Fig. \ref{fig:spectra}.
In the paramagnetic temperature regime a minimum due to a quadrupole splitting \textit{QS} was observed.
With increasing pressure the minimum is shifted to longer times showing a reduction of \textit{QS}.
In the AFM temperature regime an oscillation pattern was observed showing the presence of static magnetic order.
With increasing pressure the oscillation frequency is reduced showing a reduction of the magnetic hyperfine field \bhf.

\begin{figure}[htbp]
	\centering
		\includegraphics{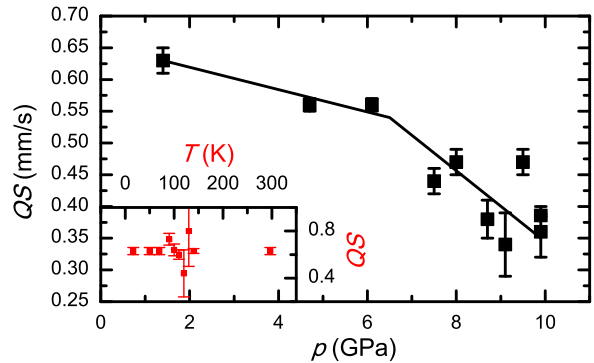}
	\caption{Pressure dependence of the quadrupole splitting \textit{QS}. \textit{QS} was taken in the paramagnetic temperature regime. \textit{QS} is temperature independent within error bars. This is exemplarily shown in the inset for 1.4\,{}GPa. Lines are guide to the eye only.}
	\label{fig:QS-p}
\end{figure}

The pressure dependence of \textit{QS}, which is directly proportional to the electric field gradient (EFG) at the Fe nucleus, is shown in Fig. \ref{fig:QS-p}.
\textit{QS} is temperature independent in the investigated temperature regime below 180\,{}K within error bars.
Therefore, \textit{QS} does not change at the magnetic phase transition which is in contrast to the FeAs-based compounds where \textit{QS} increases \cite{PhysRevB.98.014517, PhysRevB.92.134511, 1367-2630-11-2-025014, PhysRevB.83.134410}.
This indicates the absence of a charge redistribution due to the magnetic order which is most likely a result of the stronger localization of the electrons in the Mott insulator BaFe$_2$S$_3$ compared to the more itinerant FeAs-based systems.
\textit{QS} shows a monotonic reduction from $\sim$ 0.60 to $\sim$ 0.35\,mm/s at 1.4 and 9.9\,GPa, respectively.
Between 1.4 and 6.1\,GPa \textit{QS} decreases with $-$0.014(1)\,mm/s / GPa which changes to $-$0.061(9)\,mm/s / GPa between 6.1 and 9.9\,GPa.
Room temperature x-ray diffraction (XRD) experiments observed a continuous reduction of the \textit{a}-/\textit{b}-/\textit{c} lattice parameters with increasing pressure \cite{1361-6668}.
This continuous reduction would result in a constant slope of \textit{QS} as a function of \textit{p} while a change in slope indicates a change in either the electronic and/or crystallographic structure.
This change in the slope above 6.1\,GPa coincides with a discontinuous change in the S heights from the iron plane between 6 and 8\,GPa \cite{1361-6668}.
This is consistent with the fact that the EFG is very sensitive to the anion height \cite{1367-2630-11-3-035002}.
Another possible cause is an increased delocalization of the formerly localized electrons \cite{nishihara1979mossbauer}.
This might indicate that the fingerprint of the metallic state is already visible at pressures lower than $p_c$.

\begin{figure}[htbp]
	\centering
		\includegraphics{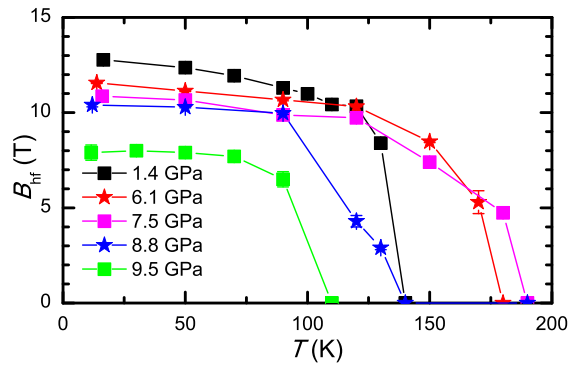}
	\caption{Temperature dependence of the magnetic hyperfine field \bhf \,{} for representative pressures. Lines are guide to the eye only.}
	\label{fig:B-T}
\end{figure}

Analyzing the temperature dependence of the magnetic hyperfine field \bhf(\textit{T}), which is shown in Fig. \ref{fig:B-T}, in the vicinity of the magnetic phase transition using an order parameter function of the type \bhf$(T)$ = \bhf$(0) \times (1 - T/T_{\mathrm{N}})^{\beta}$ the critical exponents $\beta$ were obtained.
$\beta$ increases from 0.09(1) to 0.28(2) for 1.4 and 7.5\,GPa, respectively. 
This increase suggests that the inter-ladder interaction is enhanced with increasing pressure due to a reduction of the inter-ladder Fe distances, strengthening the three-dimensional character of the magnetic order.

\begin{figure}[htbp]
	\centering
		\includegraphics{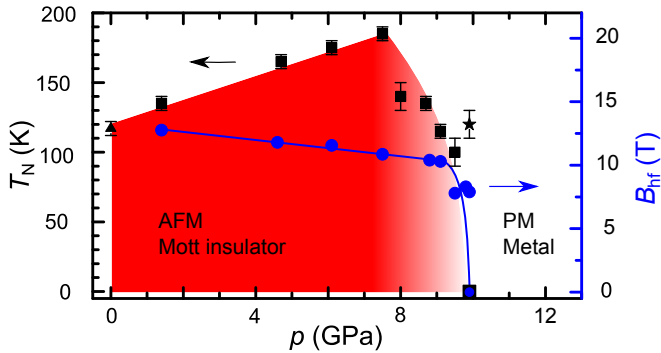}
	\caption{Magnetic phase diagram of BaFe$_2$S$_3$ as a function of pressure obtained by synchrotron M\"ossbauer spectroscopy. \tn \,{}(black square/star/triangle, left axis) and the low-temperature magnetic hyperfine field (blue circle, right axis) are plotted. The star data point shows phase separation with $\sim$ 72\% PM signal fraction. The triangle data point at ambient pressure was obtained by magnetization measurements on single crystals of the same batch. \tn \,{} continuously increases to 185(5)\,{}K at 7.5\,{}GPa  followed by a rapid suppression to zero at 9.9\,{}GPa. In contrast, \bhf \,{} is continuously reduced to 10.3(1)\,{}T at 9.1\,{}GPa followed by an abrupt suppression to zero at 9.9\,{}GPa.}
	\label{fig:T-B-p}
\end{figure}

The pressure dependence of the magnetic phase transition temperature \tn \,{} is shown in Fig. \ref{fig:T-B-p}.
Magnetization measurements on single crystals from the same batch (not shown) at ambient pressure revealed a magnetic phase transition temperature of \tn = 117(5)\,{}K.
\tn \,{} continuously increases to 185(5)\,{}K at 7.5\,{}GPa  followed by a rapid suppression to zero at 9.9\,{}GPa.
This increase in \tn \,{} is consistent with muon spin relaxation experiments conducted in the low pressure region \cite{2018arXiv180710703Z}.
Yamauchi \textit{et al}. conducted dc resistivity measurements and observed a maximum in \tn \,{} at $\sim$ 3\,{}GPa followed by a monotonic reduction with increasing pressure \cite{PhysRevLett.115.246402}.
Reasons for the discrepancy could be anisotropies in the resistivity or slightly different stochiometries \cite{PhysRevB.92.205109}.
The increase of \tn \,{} between 0 and 7.5\,{}GPa can be explained by an increase of the exchange interaction \cite{BLOCH1966881}.
The application of pressure compresses the unit cell volume and reduces the Fe-Fe distances \cite{1361-6668}.
As a consequence the exchange interaction between the Fe-atoms, and thus \tn \, may increase.

The pressure dependence of the low-temperature magnetic hyperfine field \bhf \,{} is shown in Fig. \ref{fig:T-B-p}.
\bhf \,{} continuously decreases from 12.8(1) to 10.3(1)\,{}T at 1.4 and 9.1\,{}GPa, respectively, followed by an abrupt suppression to zero at 9.9\,{}GPa indicating a first-order phase transition.
We conducted two measurements at 9.9\,{}GPa and one shows paramagnetic (PM) behavior down to 10.3\,{}K while in the second measurement a phase separation of $\sim$ 72\% PM and 28\% magnetic volume fraction with a magnetic hyperfine field of $\sim$ 8\,{}T was observed.
Taken into account that small pressure differences might occur, this result supports the first-order phase transition interpretation.
An extrapolation of our data to ambient pressure yields \bhf = 13.4(1)\,{}T at low temperatures, in agreement with a previously reported value of  $\sim$ 13.5\,{}T \cite{nmat4351} that corresponded to an ordered moment of  1.2\,{}$\mu_{\mathrm{B}}$, as determined by neutron scattering.
Other reported values for the ordered magnetic moment range between 1\,{}$\mu_{\mathrm{B}}$ \cite{PhysRevLett.117.047003} and 1.3\,{}$\mu_{\mathrm{B}}$ \cite{2018arXiv180710703Z}.
However, for the entire pressure range studied in this work, \bhf \, is in good qualitative agreement with DFT calculations that show a reduction of the magnetic moment by $\sim$ 20\% between 0 and 8\,{}GPa followed by a suppression to zero for pressures above 10\,{}GPa \cite{2018arXiv180710703Z, PhysRevB.92.085116, PhysRevB.95.115154}.

The pressure dependence of \bhf \, is in contrast to the observed increase of \tn.
XRD measurements at room temperature have shown that the Fe-S distance decreases with increasing pressure \cite{1361-6668}.
This will most likely enhance the Fe 3\textit{d}-S 3\textit{p} hybridization and thus decrease the ordered Fe magnetic moment similarly to the FeAs-based compounds \cite{PhysRevB.67.155421,PhysRevB.78.024521, PhysRevLett.101.126401}.
Moreover, DFT calculations have shown that a shorter Fe-S distance weakens the magnetic moment \cite{2018arXiv180710703Z}.
However, the strong suppression between 9.1 and 9.9\,{}GPa indicates that the Fe 3\textit{d}-S 3\textit{p} hybridization cannot be the only reason for the reduction of \bhf.

The first-order phase transition of \bhf \,{} can be explained by an increase of the electronic band width \textit{W}, as the insulator-metal transition in BaFe$_2$S$_3$ was identified to be bandwidth-controlled \cite{nmat4351}.
Calculations have shown that \textit{W} increases by $\sim$ 25\% between 0 and 12.4\,{}GPa while the Coulomb repulsion \textit{U} decreases by 6--7\% \cite{PhysRevB.92.054515}.
Therefore, by increasing pressure, \textit{W}/\textit{U} increases, giving rise to an abrupt insulator-metal transition at the critical pressure, and reducing the magnetic moment and thus \bhf \,{} to zero.

In summary, we have studied the local magnetic phase diagram of BaFe$_2$S$_3$ as a function of pressure up to 9.9\,GPa by means of synchrotron M\"ossbauer spectroscopy.
We observed upon increasing pressure a continuous increase of the magnetic ordering temperature, that reaches a maximum value of 185(5)\,{}K at 7.5\,{}GPa, and decreases abruptly to zero at 9.9\,{}GPa.
In contrast, the low-temperature hyperfine field decreases by $\sim$ 20\% between 1.4 and 9.1\,{}GPa, followed by a sharp suppression to zero at 9.9\,{}GPa indicating first-order phase transition.
Additionally, the magnetic order appears to acquire an increasingly three-dimensional character between 1.4 and 7.5\,{}GPa, in line with an increase of the inter-ladder interaction.
As a result of this study, the pressure dependence of the magnetic order in BaFe$_2$S$_3$ can be qualitatively explained by a combination of a bandwidth-controlled insulator-metal transition, a pressure enhanced exchange interaction between Fe-atoms, and Fe 3\textit{d}-S 3\textit{p} hybridization.

\textbf{Acknowledgments}
\vspace{12pt}
\\
Part of this work was funded by the Deutsche Forschungsgemeinschaft (DFG) -- MA 7362/1-1 and BU 887/15-1.
M.L.A. acknowledges support from the DFG within GRK 1621.
This research used resources of the Advanced Photon Source, a U.S. Department of Energy (DOE) Office of Science User Facility operated for the DOE Office of Science by Argonne National Laboratory under Contract No. DE-AC02-06CH11357.
Support from COMPRES under NSF Cooperative Agreement EAR-1606856 is acknowledged for partial support of W. Bi.

\bibliography{BaFe2S3}
\end{document}